# Packaged Software Implementation Requirements Engineering by Small Software Enterprises

Issam Jebreen, Robert Wellington and Stephen G. MacDonell
SERL, School of Computing and Mathematical Sciences
AUT University
Auckland 1142, New Zealand
issam.jebreen@aut.ac.nz, rwelling@aut.ac.nz, stephen.macdonell@aut.ac.nz

**Abstract**

*Small to medium sized business enterprises (SMEs) generally thrive because they have successfully done something unique within a niche market. For this reason, SMEs may seek to protect their competitive advantage by avoiding any standardization encouraged by the use of packaged software (PS). Packaged software implementation at SMEs therefore presents challenges relating to how best to respond to misfits between the functionality offered by the packaged software and each SME's business needs. An important question relates to which processes small software enterprises – or Small to Medium-Sized Software Development Companies (SMSSDCs) – apply in order to identify and then deal with these misfits. To explore the processes of packaged software (PS) implementation, an ethnographic study was conducted to gain in-depth insights into the roles played by analysts in two SMSSDCs. The purpose of the study was to understand PS implementation in terms of requirements engineering (or 'PSIRE'). Data collected during the ethnographic study were analyzed using an inductive approach. Based on our analysis of the cases we constructed a theoretical model explaining the requirements engineering process for PS implementation, and named it the PSIRE Parallel Star Model. The Parallel Star Model shows that during PSIRE, more than one RE process can be carried out at the same time. The Parallel Star Model has few constraints, because not only can processes be carried out in parallel, but they do not always have to be followed in a particular order. This paper therefore offers a novel investigation and explanation of RE practices for packaged software implementation, approaching the phenomenon from the viewpoint of the analysts, and offers the first extensive study of packaged software implementation RE (PSIRE) in SMSSDCs.*

**Keywords**: *requirements engineering; packaged software implementation; ERP; SMEs*

## 1. INTRODUCTION

In recent years the market through which packaged software (PS) is sold to large companies has become saturated [1]. PS companies and vendors have therefore begun to target small to medium-sized business enterprises (SMEs), and various midrange and less complex software packages have been developed for this purpose [2]. SMEs are of critical importance to many economies; according to Snider et al. [24] SMEs "with less than 500 employees provided 51 per cent of all employment in the USA as of March, 2004 and 64 per cent of all Canadian private sector employment in 2005. In the European Union, firms with 250 employees or less provided 67 per cent of employment in 2003". While SMEs are an integral part of economies, they face specific challenges when implementing packaged software [2, 24, 25].

SMEs are considered to be fundamentally different from large enterprises in several respects [3]. Some distinguishing characteristics of SMEs include ownership type, culture, structure, and market orientation [5, 6]. Researchers have found that when it comes to IT/IS adoption, SMEs are constrained by limited resources and limited IS knowledge, or by a lack of IT expertise [3, 4]. These distinguishing characteristics may influence the PS implementation issues SMEs face [2] and lead to PS implementation being a challenge for many SMEs [2, 7, 8]. Studies of PS implementations have argued that findings relating to implementations in large companies cannot be applied to SMEs [3, 4]. Despite the importance of PS implementation being recognized by these former studies, there has been little research exploring this issue further. In particular, discussions about SMEs rarely occur in the PS implementation literature, and the question of whether or how the structure of SMEs shapes software throughout its implementation life cycle is rarely mentioned [2].

Since SMEs generally thrive because they have successfully done something unique within a niche market, they may seek to protect that competitive advantage by avoiding any standardization encouraged by the use of packaged software. Hence, the one of the key challenges of PS implementation in SMEs relates to functional misfits between the functionality offered by the packaged software and the business needs of the SME. In order to understand such challenges fully, it would be beneficial to understand the requirements engineering processes that Small to Medium Sized Software Development Companies (SMSSDCs) apply in order to identify misfits between the PS functionalities and SME business processes, in order to achieve a better fit. Gaining a better understanding of such processes is necessary as researchers have argued that most current requirements engineering practices are unsuitable for SMSSDCs [10, 11].

This study addresses the lack of understanding of Packaged Software Implementation (PSI) in terms of Requirements Engineering (RE) by Small to Medium Sized Software Development Companies (SMSSDCs). It investigates this phenomenon from the perspective of the SMSSDCs. RE as it relates to Packaged Software Implementation will hereafter be referred to as PSIRE.

In focusing on developing an understanding of PSIRE we construct a "theory of understanding" to represent how and why events occur during the implementation process of PS in terms of RE. According to Gregor [12] this type of theory is suitable when the researcher uses an interpretive paradigm.

The rest of this paper is organized as follows: in Section II we provide a review of previous literature relevant to our study; in Section III we briefly describe the research method; in Section IV we present our findings and results, which are then discussed in Section V; Section VI delivers our conclusion and considers future work.

## 2. LITERATURE REVIEW

The poor use of requirements engineering (RE) practices has often been identified as one of the major factors that can jeopardize the success of a software project [13, 14], and, conversely, following appropriate RE practices has been found to contribute to the success of software projects [15, 16]. Aranda et al. [14] state that gathering and managing requirements properly are key factors when it comes to the success of a software project. However, it is not possible to improve RE practices until areas that need have been identified [11, 17], and the solution for improving RE practices will be different in each company; a one-size-fits-all approach does not work in such a scenario [11, 14, 17].

In general, SMSSDCs are unable to apply conventional RE methods and techniques without modification [14, 16, 17]. In addition, shortcomings in applying RE methods may arise due to time constraints [16]. Bürsner & Merten [17] note that "RE research has to intensify the investigation of RE practices in SMEs [SMSSDCs]. Otherwise SMEs [SMSSDCs] will have to continue their search for methodical orientation and dedicated tool support. Normally, the people responsible for requirements in SMEs are ambitious, but suffer from scarcity of resources. Their time for doing experiments and trying different methods is very limited. They need quick methodical improvement of requirements elicitation, documentation, communication and traceability as well as more continuity of requirements management through the whole software lifecycle".

Karlsson et al. [18] observe that there are "several studies that concern or include RE issues. However, none of these focus primarily on packaged software development and implementation. Furthermore, in most of these studies, the studied projects and organisations are mainly large, both in terms of the number of persons and requirements involved, and in terms of the duration of the projects". Quispe et al. [11] highlight that "there is a lack of knowledge about the requirements engineering practices in these types of companies [small-medium]". In short, researchers encounter a general lack of information about how RE is carried out in packaged software companies. It is difficult for researchers to gain knowledge about how SMSSDCs carry out RE given that most SMSSDCs seldom request external support, probably due to limited finances. However, RE research should eventually enable those companies to become aware of more state of the art or innovative RE techniques and to be able to improve their RE practice without external help [16].

Merten et al. [16] argue that SMSSDCs may not need to have formal or conventional forms of RE. Instead, "light organization and unconventional RE" may work better for many SMSSDCs. They also discuss the various RE models that have been provided in previous studies; for example, Olsson et al. [19] presented a pragmatic framework for RE in SMSSDCs. However, Merten et al. [16] suggest that the list provided needs to be expanded in future because the selection of RE techniques is a central problem in all aspects of process improvement. They note another study by Hardiman [20] but observe that the RE practices and techniques discussed in that study seem to be tailored toward particular individual SMSSDCs and therefore do not seem to offer solutions that can be applied to the whole domain. Pino et al. [21] provide an extensive list of Software Processes Improvement (SPI) models, and discuss methods based on ideas put forward by the Software Engineering Institute (SEI) or by the International Organization for Standardization (ISO). However, Pino et al. [21] note that many of the models proposed by these two organizations could be too complex for SMSSDCs to apply.

This study presumes that the specific characteristics of SMEs, SMSSDCs, and PS implementation may influence RE, while recognizing that recent literature has paid little attention to RE as it applies to PS implementation from the perspective of SMSSDCs [2, 10]. While software engineering comprises a group of influential approaches that are often considered good practice, including 'structured programming', 'stepwise refinement', and collecting 'a complete set of test cases' [23], these approaches do not apply for PS implementation. Dittrich et al. [23] argue that such implementation requires "independent consideration".

In this study, we set out to discover which RE practices are actually being used by SMSSDCs during packaged software implementation. We also highlight some of the dynamics and complexity that these SMSSDCs face, as well as their reaction to these challenges. Putting the organisation and organisational process at the centre of attention, this research advances our understanding of packaged software implementation from a work organisation point of view, and in terms of requirements engineering. The SMSSDC is the point of entry for this study, and this research provides a SMSSDC's perspective.

## 3. RESEARCH APPROACH

Ethnographic research was conducted over a period of seven months in an SMSSDC. Data were collected throughout the field work. The three data collection methods, namely, interviews, participant observation, and focus groups, were used due to their suitability for qualitative research.

### A. Research setting

The software development company that participated in this research was established in 1997. It has 40 employees, working in marketing and sales, as analysts, developers, and in management teams. The services they offer include software development, systems integration, and software localization. The company's software products deal with accounting, inventory management, purchasing, retail, school management, freight management, and human resource management.

The total number of individual participants was 16, comprising a mix of analysts, developers and team leaders. The majority of the participants had between 3-10 years experience in the field. Most of the participants had experience working as analysts, designers, and developers at the same time, with business application software and database system software. During the seven month period the first author observed 35 project cases. The specific form of PS considered was Human Resource (HR) software, Enterprise Resource Planning (ERP) software, Special Solution software, such as a school management system, Restaurant Management software, and Point of Sale (PoS) software. Figure 1 visually represents the number of cases observed.

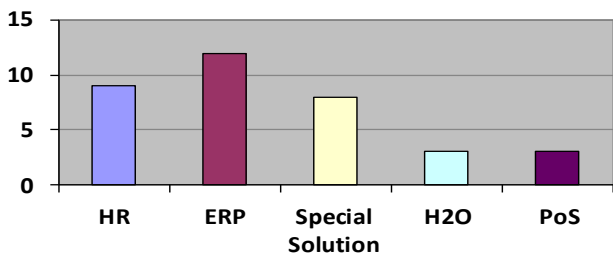

Fig. 1. Number of cases observed

### B. Data analysis

Inductive analysis, as used in this study, refers to an approach that primarily uses detailed reading of raw data to derive concepts, themes, and models through the researcher's interpretations of the raw data [27, 29]. During ethnographic research the ethnographer goes through a learning process at the same time as conducting their research. This process allows any prior presumptions that are found to be false to be redefined, reformed, or discarded [27]. The researcher is then more open to experiencing what is going on around them, to paying attention to the details of the process, and to observing what was actually happening in the company, rather than trying to search for relevant data.

In this study, an initial round of field observations was conducted to find interesting topics associated with the company's practices for PS implementation. We wished to discover the situations in which PS implementation occurs and understand the process that participants apply. After the initial field work, an initial round of coding was conducted in order to single out descriptive and interpretive codes [29].

Excerpts from the text of the interview/focus group transcripts or specific phrases from the transcripts were assigned as interpretive codes in the initial coding. Further analysis of the initial codes supported the grouping of similar descriptive and interpretive codes in order to form categories with common themes. The theme names were derived based on the concepts in the organizations' projects. This concluded the second round of coding. For example, 'Present software', 'Explain software functions', 'Help users', and 'Users' business Support' were categorized under software demonstration (see Table I). A third round of analysis was conducted to derive higher-level concepts that would comprise the theoretical constructs in a model of the PSIRE. Engaging in a third round of analysis aids the researcher in reaching a higher level of abstraction [29]. For example, 'live scenario software demonstration' was conceptualized as a strategy that was intended to help analysts convince clients about the software solution, as they demonstrate and negotiate a possible solution. Theoretical concepts of themes are identified by an abstraction that describes the themes.

TABLE I. CODING PROCESS

| Data Extract | Codes for |
|---|---|
| Question: Tell about software demonstration? The software we present is based on the notes from the sales team about the user's interest in potential software. Then we present the functions of the software that supports the user's business….. I think that helps the user know what their expectations could be for the software functions | Present software Sales team report User's Interest Explain software functions Users' business Support Help users Users' expectations |
| It is good for us to make a software demonstration, in which we start to present a possible solution for users' issues. The flexibility that we want to have during software demonstration was constrained by a time limit since we only have one hour and a half to present our software….so we have to do our best to explain our software functions to the users. | Benefits of software demonstration Present a possible solution Users' Issues Constraints Time limitation Present software Explain software functions |

## 4. FINDINGS AND RESULTS

Our inductive analysis of the collected data, across participants, provided a rich set of findings to inform a view of PSIRE. Five main processes that emerged from the analysis are: (1) PS feasibility study; (2) installation; (3) PS software demonstration; (4) identify misalignments; and (5) assessment.

## A. Feasibility Study

The feasibility study (FD) stage in PSIRE resembles, at a high level, such feasibility studies as those used in traditional RE. This is because feasibility studies in traditional RE and the FD stage discussed here are similar in terms of their purpose, such as dealing with software objectives, time and budget. However, at the more detailed practical level, the feasibility study (FD) stage in PSIRE practice has its own characteristics.

The results indicate that the analysts attempted to define the scope of the packaged software they were implementing during discussions with potential clients about their needs. Analysts believed it was important to carry out this scoping process early on since this would help them to construct a sensible software offer, and help all involved to exercise control over the time taken for implementation. Analysts felt that collecting such information not only provided them with details about what the new software needed to do, but also helped them to see what its limitations would be and what features or modifications would be unnecessary. This therefore helped them to design a PS offer that would likely suit the client.

Software scoping was generally limited to finding out information about the core requirements of the system or solution to be designed. This was found to principally involve transaction issues and output format issues; during the scoping process the analysts were not concerned with discovering detailed requirements. Another aspect that was considered during the scoping process was the cost of implementing the software. Analysts needed to take into account what clients might be prepared to pay and what software was worthwhile for their own company to implement.

The study results indicate many of the aspects that analysts took into consideration when making a software offer, and pre-conditions to be met by packaged software that are mentioned in the software offer. When creating a packaged software offer, the SMSSDC decided on the scope of the offer and exactly how to develop the software based on the client's initial requirements, the modifications requested by the client, the extent of the modifications, and the technical requirements involved in meeting such requirements.

It was also found that the software company applied different criteria of assessment when considering how to make a software offer and when estimating the effort and time needed to develop, customize, and modify the packaged software. These assessment criteria related to various offer elements. The assessment criteria mentioned by the General Manager related to: New Features, Customization, Input/Output, and Technical Issues.

## B. Installation

The determination of users' needs consists of identifying misalignments by conducting discussions with users to determine what feature wants and needs the user has in relation to the packaged software on offer. The analysts observed in this study commonly installed a copy of the packaged software in order to identify misalignments between the packaged software's technical requirements and the user's IT infrastructure. Analysts then used the installed copy of the software to provide a software demonstration to users. This helped the analysts to identify the business misalignments between the software functionalities and the user's business process functions, and to gather information about necessary customization, new features, and output requirements.

In one of the cases observed, a Human Resource Management System (HRMS) had been offered to an organization. Since the client accepted the software offer, further in-depth understanding of the users' needs was required. The analysts started to identify technical misalignments between the packaged software's technical requirements and the user's IT infrastructure via the installation of a copy of the packaged software. This was done to determine if there would be any software integration issues or software infrastructure issues involved with a full implementation of the software. When implementing packaged software, there is a need for certainty regarding whether the infrastructure required by the packaged software is in place at the client's site. Several technical issues were discovered by installing the copy of the software. For example, issues were found that related to server compatibility. Other issues were found on the users' desktop side, such as their computer missing some components that were related to running files.

Hence, it is clear that analysts need to identify the misalignments between the software's technical requirements and the users' infrastructure capability in order for the PS to be implemented. However, how is such a copy of the software used by analysts to identify business misalignments: new features, customizations, and output adjustments?

## C. Software Demonstration

The analysts spoke about this installation of the copy as a way to educate users about the software's functionalities and to increase users' participation in discussions. After the software was installed successfully, analysts used the version of the software to carry on identifying misalignments. For example, in the case of this HRMS software, the users' issues were categorized under transaction issues such as 'add employees', 'bonuses mechanism', and 'payments made for uniforms', and output format issues. Analysts spoke of how using this installed copy of the software could minimize the customization effort.

A transaction misalignment was found that required customization of the software. Analysts explained to the users the functionality related to payments made for uniforms. The users accepted the interface layout and the output data but asked for the customization of a relationship between 'payments made for uniforms' and 'employees' salary'. That is, the software needed to include a mechanism by which a fee was deducted from the first month of an employee's salary, as a guarantee for uniforms, and then returned to the employee after three months. In this case, the analysts minimized the customization effort by explaining how the software could help users when kept in its present form, and then agreeing to customize the software in terms of the transaction formula.

There was strong consensus amongst the analysts interviewed that analysts should consider carrying out software demonstrations for packaged software as a means of convincing users that there were often alternative solutions to misalignments. The general recommendation from analysts who participated in this study was that demonstrations of a trial version of the packaged software should be used as part of the implementation process to educate users about the software's functionalities, to increase users' participation in discussions, and to discover and discuss user needs and misalignments.

More comprehensive discussions of misalignment types can be found in Yen et al. [30] and Sia & Soh [31]. However, these approaches do not hold for the investigated small – medium sized software development companies as their studies consider only the perspective of users, not the perspective of the packaged software companies.

### D. Identify misalignments

In general terms, analysts respond to the discovery of a misalignment in one of two ways: either that the user's company should adopt the packaged software and its functionality as is – a decision that might require the company to change their business processes – or that the packaged software needs modification in terms of customizing a function or adding new functionality. However, various factors need to be taken into consideration before the analyst decides what action to take in response to finding a misalignment.

The analysts first need to determine whether the misalignment that has been discovered is in fact an 'actual' misalignment, or only a perceived one (a distinction that does not exist in the RE practice for bespoke software). A misalignment is 'actual' only when the software does not support such a transaction or does not support a specific transaction formula. A misalignment might seem to exist in cases where the software functions actually support a particular desired transaction, but in a different way than expected.

If an 'actual' misalignment is found, the impact of various factors related to the misalignment must be considered. These factors may make it impossible to fix the misalignment. For example, analysts need to determine whether the misalignment is within the software scope or outside the software scope, and will also need to consider the size of the user's organization. After carrying out such assessments of misalignments, the analysts can choose a course of action.

However, the individual SMSSDC's strategy for dealing with users' needs and misalignments must also be kept in mind. For example, the SMSSDC observed in this study wished to minimize customization of software as much as possible. Meanwhile, one last factor bears some importance when misalignments are discovered: finding a misalignment is not necessarily negative, and may in fact hold some benefits. The presence of a misalignment may sometimes provide the opportunity for SMSSDC developers to test aspects of their software or to improve their current packaged software.

### E. Assessment

Assessment generally involves making decisions related to misalignments that have been found between the packaged software and the client's requirements or the client's business environment. The misalignments found may relate to output functions and to the user interface, but more commonly relate to transactions. While engaging in 'assessment', analysts need to consider both the software dimension and business dimension of responding to misalignments. In terms of the software dimension, there could be risks to the software if modifications are made. In terms of the business dimension, the analysts will consider whether dealing with the misalignment is within their work domain, and whether there is any benefit to their organization from dealing with the misalignment.

#### 1) Identifying 'actual' & 'perceived' misalignments

As suggested above, before assigning misalignments to a specific category such as 'new feature', 'customization', and 'output', it should be decided whether a particular misalignment is 'actual' or only 'perceived'. A misalignment might be perceived to exist in cases where the software functions actually support a particular desired transaction, but in a different order. One such example occurred in the case relating to HRMS software, when the accounting manager asked the analysts to add some attributes to employees' salary reports. However, the analysts explained that these attributes were already represented by other reports. As a result, the accounting manager accepted the software report order as it was, without requesting further changes. However, in another case, as mentioned above, analysts explained to the users the functionality related to payments made for uniforms. The users accepted the interface layout and the output data but asked for the customization of a relationship between 'payments made for uniforms' and 'employees' salary' that would involve a transaction formula that was not supported by the software.

Both of these situations involve misalignments, but in the case where the accounting manager asked for attributes to be added to the salary reports, the misalignment can be categorized as a 'perceived' misalignment since it could be 'worked around' by carrying out a process in a slightly different way than was initially desired (by finding the attributes in other reports). However, the misalignment that was found in relation to payments for uniforms can be categorized as an 'actual' misalignment because the misalignment was such that the user's business process could not work unless a customization was made. As sometimes happens with RE for bespoke software, in packaged software implementation, analysts may use work-arounds, but this is in order to minimize customization, rather than to reduce conflicts between requirements.

#### 2) Minimizing customization

The general managers and analysts spoken to during the field work for this study supported the idea that users should adopt a package's software functions as they are and change their own business processes, rather than seek to modify the software. One reason for this recommendation was that the business processes of the user may be so idiosyncratic that carrying out the modifications desired by the user might have significant impact on the software functions. The mechanism of minimizing customization and dealing with misalignments involves gaining an understanding of what is redundant in software for a particular user context, what functions are essential for the operation of the software, and which customization requests can be met without disrupting the software. Such considerations must extend to involve users' needs, the scope of the project, and customization risk. In packaged software implementation, when users inform the analyst that a particular function is redundant to their needs, the analyst has to consider whether the unwanted function is actually connected to other functions of the software and whether changes made to the redundant function may impact other areas of the software. In the case that the unwanted function cannot be deleted, the user needs to adopt the functions of the software, even if it does not match the user's current business processes.

During this study, however, it was found that despite the potential problems associated with customization and even though analysts try to encourage users to adopt the existing functions of the software package, when there are 'actual' misalignments analysts are left with no choice but to agree to customization.

*3) Organisation size and price of software*

The level of customization engaged in by a company often corresponds to company size. Some clients, especially small and medium sized enterprises (SMEs), may wish to carry out a large degree of customization. This is because many such companies view it as essential that they keep the 'best practices' that they believe give them a competitive advantage. A larger SME may be more able to afford a greater number or extent of customizations and to be able to match the software price. During this study, it was suggested by analysts at the SMSSDC that if an organisation can afford to pay for a particular customization, it will likely go ahead. Customization decisions are usually associated with challenges involving increased costs and longer implementation periods. However, the existence of actual misalignments, factors related to the user's organisation size, and the price that the client pays for the software all have an influence on customization decisions.

*4) Benefits derived from users' needs/misalignments*

A SMSSDC may sometimes benefit from the identification of misalignments. For example, when a user points out functionalities that the package does not provide and that they desire, even if the SMSSDC cannot develop the new function in time to include it in the current package or considers it too risky or costly to include in the current package, the SMSSDC may still have been provided with an idea for a useful function to add to a to the next release of the package. It may therefore be the case that unforeseen benefits can be derived from discovering misalignments.

## 5. DISCUSSION

Previous studies have typically focused on the users' perspective on PS implementation, revolving around their attempts to select packaged software that fits their process. By changing the perspective used from 'outside' the SMSSDC to 'inside' the SMSSDC, this research provides a new understanding of an RE process for packaged software implementation for SMSSDCs (a process that we term 'PSIRE'). This process should help to identify misalignments between users' needs and packaged software functionalities. Part of our new understanding of this process involves our proposing a Parallel Star Model for PSIRE which is based on empirical observations of analysts made during our ethnographic research.

The Parallel Star Model is especially designed to support the parallel processes (that include feasibility study, assessment, implementation, software demonstration, and identifying misalignments) of PSIRE, especially as conducted at SMSSDCs. It bears both similarities to and differences from the star model theorized by Hartson & Hix [32]. The first similarity between Hartson & Hix's [32] model and our own lies in the use of a star-shaped configuration to show the possible interconnections between different processes involved in the development and provision of packaged software (or in Hartson & Hix's study, the development of interfaces). Both models feature a group of processes that are connected to each other by means of a central step which relates to making assessments about the next action to take or activity to engage in. In Hartson & Hix's [32] model, this central step is referred to as 'usability evaluation'; in our model, the central step is 'assessment'. In abstract terms, these central terms are very similar, involving pausing to check information and to carefully consider the next step. In more practical and specific terms, the central steps differ. Hartson & Hix's [32] model focuses on human-computer interface development and the central issue involved in their 'evaluation' step is to address the interface usability. The forms of 'assessment' involved in our Parallel Star Model for PSIRE (shown in Figure 2) are quite different, as PSIRE has different concerns. In our Parallel Star Model, 'assessment' generally involves making decisions related to misalignments that have been found between the packaged software and the client's requirements or the client's business environment. The misalignments found may relate to output functions and to the user interface, but more commonly relate to transactions. While engaging in 'assessment', analysts need to consider both the software dimension and business dimension of responding to misalignments. In terms of the software dimension, there could be risks to the software if modifications are made. In terms of the business dimension, the analysts will consider whether dealing with the misalignment is within their work domain, and whether there is any likely benefit to their organisation.

Apart from the fact that PSIRE involves different concerns than those that are relevant to human-computer interface development, the Parallel Star Model contains one large structural change from Hartson & Hix's model. While their model showed that there can be flexible connections between different processes, their model does not show any processes running in parallel. The Parallel Star Model, however, shows not only that the processes involved in PSIRE are interconnected in flexible ways, arranged around the central step of 'assessment', but that multiple PSIRE processes can be followed simultaneously.

All of these considerations lead to a different kind of RE life cycle approach in the Parallel Star Model (see Figure 2). This model shows that during PSIRE, processes can be carried out in parallel and do not always have to be followed in a particular order. There are very few constraints to the sequence in which processes can be followed. This model shows that analysts can theoretically carry out multiple processes at the same time (hence, it is a parallel process model).

Because multiple processes can be carried out at the same time or swapped between quite easily, one particular benefit of the model is that it reduces the ordering constraints acting upon process activities. For example, analysts do not necessarily need to have found all of the misalignments that are present before working on the training of users. In fact, analysts can train users in how to use the software at the same time as they identify misalignments. They can also validate that they have changed the software to deal with a misalignment at the same time as training the users, or can identify technical misalignments in parallel with software installation. Analysts can also move back and forth between finding misalignments and developing a solution to the misalignments, and looking for more misalignments. It is essential to support continual assessment and iteration during PSIRE, including smaller loops of iteration; the Parallel Star Model supports such an approach.

The feasibility study and the installation process are the initial processes needed to set up the software environment. This is the initial constraint on the model; these two actions must be taken before analysts can use this model. Once the software environment is set up, and the model is entered, there is only one major constraint on the model and the order in which steps are taken: this is that analysts should go through the central 'assessment' process before moving on to the next activity. When analysts are engaged in the central assessment process, the results of each process that has been carried out are assessed. This is done in order to help analysts make decisions on how to next take action. In general, a different kind of assessment is required after each different process in this model. Figure 2 shows that there are three different kinds of 'assessment criteria'. For example, in terms of deciding on a project's feasibility, analysts will not be able to take their next step without addressing how the client's organisation structure should affect their decision, or how their own work domain might impact their decision.

As shown in Figure 2, the form of assessment focusing on the software dimension of implementing packaged software in response to misalignments involves addressing the risk of adding new features, the risk of customization, the output customization risk, and the technical needs of dealing with the misalignment. Analysts will consider whether they can or should carry out all of the modifications desired by the user, and what technical risks or risks to the software would be involved in carrying out such modifications. They will assess, for example, whether the changes made would have significant impact on the software functions, and especially whether they would disrupt essential functions. They will also assess whether the software may be disrupted even if a non-essential ('redundant') function is modified or removed.

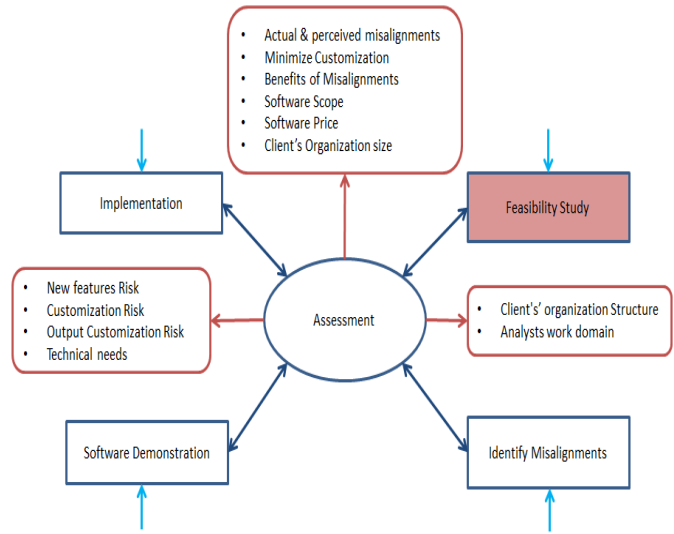

Fig. 2. PSIRE Parallel Star Model

The form of assessment dealing with the business dimension of making changes to the software involves addressing whether misalignments are actual or perceived, making assessments related to the preference to minimize customization, considering the client organisation's size, considering the software scope and the software price, and addressing the possible benefits to be gained from working with misalignments. The first consideration they make is whether a misalignment that has been discovered is an 'actual' misalignment, or only a perceived one. Even when the misalignment that has been found is 'actual', analysts will still stop to determine whether the misalignment is within or beyond the software scope. Here, 'scope' is determined by looking back at the original software offer that the analyst company made to the clients during the pre-implementation. The size of the client's organisation and the price they are willing to pay for software or for customizations are also considered during 'assessment'. When considering whether to go ahead with customizations, analysts consider the size of the user's organisation, because larger organisations can generally better afford customizations.

The Parallel Star Model is flexible, as there are very few constraints involved: the only major limitation on analysts is that they will usually need to go through the central assessment process before moving on to beginning a new process. However, this step of engaging in assessment is only needed if the analyst actually needs to consider the risks involved with making a particular decision or engaging in a particular action. If there are no risks associated with a particular decision or action, the analyst can omit the assessment step and move on to the next desired process.

## 6. CONCLUSION

Given the growing importance of packaged software it is increasingly necessary to understand the engineering practices associated with packaged software implementation. This study reports an in-depth, ethnographic investigation of requirements engineering practices for implementing packaged software at SMEs, but does so from the point of view of analysts working for SMSSDCs. It also captures the processes involved in RE for PSI by proposing and describing a Parallel Star Model that supports the parallel processes of PSIRE.

Some limitations to this study should be acknowledged: the results that we have gained during this study could be validated further if researchers gained data about PSIRE from other SMSSDCs. As suggested in our 'Discussion' section, it could be desirable for researchers to shift their focus from examining users' organisations to examining the views SMSSDCs have of packaged software implementation. If such action is taken, researchers and practitioners will be able to gain a more complete understanding of all of the sites and participants involved with packaged software implementation.

Another topic related to PSI that would be very interesting to investigate is how the philosophy behind release plans for packaged software differs between large packaged software development companies and SMSSDCs. From observations made here and in previous literature, it appears that large packaged software development companies tend to have very detailed release plans and schedules for future packaged software products, mapped out months or years in advance, while SMSSDCs may take a more *ad hoc* approach to release planning that instead involves continuous improvement of their product in response to clients' requirements and clients' responses to their product. Other research areas of interest would be discovering tools that could support misalignments management for SMSSDCs and developing a document template suite that could support PSIRE for SMSSDCs.